%% file: template.tex
\documentclass{article}
\usepackage{arxiv}

\usepackage[utf8]{inputenc} 
\usepackage[T1]{fontenc}    
\usepackage{hyperref}       
\usepackage{url}            
\usepackage{booktabs}       
\usepackage{amsfonts}       
\usepackage{nicefrac}       
\usepackage{microtype}      
\usepackage{lipsum}		
\usepackage{graphicx}
\usepackage{doi}
\usepackage{amsmath}
\usepackage{subcaption}
\usepackage[dvipsnames]{xcolor}

\usepackage[style=numeric]{biblatex}
\title{Data Compression for Time Series Modelling:\\A Case Study of Smart Grid Demand Forecasting}

\author{Mikkel Bue Lykkegaard \\
	Agriculture and Digitalization \\
	Danish Technological Institute \\
	Skejby, Denmark \\
	\texttt{mbly@teknologisk.dk} \\
	\And
        Svend Vendelbo Nielsen \\
	Agriculture and Digitalization \\
	Danish Technological Institute \\
	Skejby, Denmark \\
	\texttt{sven@teknologisk.dk} \\
        \And
        Akanksha Upadhyay \\
	Agriculture and Digitalization \\
	Danish Technological Institute \\
	Skejby, Denmark \\
	\texttt{akaupad@gmail.com } \\
	\And
        Mikkel Bendixen Copeland \\
	Green Energy Systems \\
	Danish Technological Institute\\
	Aarhus, Denmark \\
	\texttt{mico@teknologisk.dk} \\
	\And
        Philipp Trénell \\
	Agriculture and Digitalization \\
	Danish Technological Institute \\
	Skejby, Denmark \\
	\texttt{phtr@teknologisk.dk} \\
}

\renewcommand{\shorttitle}{Data Compression for Time Series Forecasting}

\begin{document}
\maketitle

\begin{abstract}
Efficient time series forecasting is essential for smart energy systems, enabling accurate predictions of energy demand, renewable resource availability, and grid stability. However, the growing volume of high-frequency data from sensors and IoT devices poses challenges for storage and transmission. This study explores Discrete Wavelet Transform (DWT)-based data compression as a solution to these challenges while ensuring forecasting accuracy. A case study of a seawater supply system in Hirtshals, Denmark, operating under dynamic weather, operational schedules, and seasonal trends, is used for evaluation.

Biorthogonal wavelets of varying orders were applied to compress data at different rates. Three forecasting models—Ordinary Least Squares (OLS), XGBoost, and the Time Series Dense Encoder (TiDE)—were tested to assess the impact of compression on forecasting performance. Lossy compression rates up to \( r_{\mathrm{lossy}} = 0.999 \) were analyzed, with the Normalized Mutual Information (NMI) metric quantifying the relationship between compression and information retention. Results indicate that wavelet-based compression can retain essential features for accurate forecasting when applied carefully.

XGBoost proved highly robust to compression artifacts, maintaining stable performance across diverse compression rates. In contrast, OLS demonstrated sensitivity to smooth wavelets and high compression rates, while TiDE showed some variability but remained competitive. This study highlights the potential of wavelet-based compression for scalable, efficient data management in smart energy systems without sacrificing forecasting accuracy. The findings are relevant to other fields requiring high-frequency time series forecasting, including climate modeling, water supply systems, and industrial operations. 
\end{abstract}

\keywords{Demand forecasting \and Time series modelling \and Data compression techniques}

\section{Introduction}
Time series forecasting plays a critical role in smart energy systems by enabling accurate predictions of e.g. energy demand, renewable energy generation, and grid stability. For example, forecasting electricity demand allows utilities to optimize power generation and reduce operational costs, while wind and solar power forecasting facilitates the integration of renewable energy into the grid \cite{HONG2016896}. Data-driven predictive techniques for time series have been utilitized succesfully in smart grid applications such as short-term load forecasting \cite{SMYL202075} and renewable resource prediction \cite{7926112}. Beyond energy systems, time series analysis is a fundamental tool across various scientific disciplines, including climate science, finance, and healthcare. It enables researchers and practitioners to identify trends, seasonal variations, and anomalies in temporal data. These datasets are derived from diverse sources, such as financial markets \cite{sezer2020financial}, weather monitoring systems \cite{barrera2022rainfall}, medical sensors \cite{haradal2018biosignal}, and industrial process logs.

In recent years, the proliferation of sensors and devices in the Internet of Things (IoT) has significantly increased the volume of generated time series data \cite{asghari2019internet} across application domains. Given the growing availability of time series data, efficient data compression techniques are crucial for reducing storage costs while preserving the utility of the data \cite{correa2022lossy}. Traditional subsampling methods, where e.g. the sampling frequency is simply reduced, may result in significant loss of important information. Many techniques for time series data compression have been developed, each balancing different trade-offs in compression ratio, explainability, and adaptability to changing data. Dictionary-based methods, for instance, identify recurring patterns within time series and compress them using pre-learned dictionaries. While effective for datasets with repetitive or periodic structures, their reliance on tailored dictionaries can reduce adaptability when the data distribution shifts \cite{chiarot2023survey}.

Deep learning-based compression algorithms, such as autoencoders, have demonstrated impressive performance, achieving some of the highest compression ratios in benchmarks like the Large Text Compression Benchmark \cite{mahoney2011large}. These methods excel at capturing complex patterns in data but may struggle to adapt to changes in the data, as they rely on pre-trained models that can degrade without retraining \cite{chiarot2023survey}. Additionally, their compressed representations are typically less interpretable than simpler techniques, and retraining can be computationally expensive.

Functional Approximation (FA) methods provide interpretable approaches to time series compression \cite{chiarot2023survey}. One such method is the Discrete Wavelet Transform (DWT) which decomposes time series into components across multiple frequency bands, retaining key patterns while discarding less relevant information. A key strength of the DWT is that wavelets form a broad class of approximation functions, capable of representing a wide variety of data with both local and global features \cite{mallat_wavelet_2009}. This flexibility not only improves compression, but also sometimes improves subsequent modeling tasks, as demonstrated by applications that improve the accuracy of forecasts for wind power \cite{liu2019wavelet} and electricity prices \cite{conejo2005day}.

In this paper, we evaluate the performance of different orders of biorthogonal DWTs for data compression, focusing on their impact on time-series forecasting accuracy rather than just reconstruction error. The study is applied to a seawater production and distribution system at Forskerparken, Hirtshals, which manages water intake, storage, and distribution of seawater across interconnected tanks. The operation of the system is one piece of a larger puzzle encompassing the smart operation of the power grid on the docks. The seawater distribution system exhibits strong temporal and variable correlations influenced by operational schedules, long-term seasonal trends, and weather. By assessing the effectiveness of compression algorithms in preserving the information most relevant to the forecasting task, this paper addresses a critical consideration for time series data: ensuring compression supports the actual purpose of data collection, namely accurate prediction of future behavior.

\section{Methods}
\subsection{Data}
The dataset analyzed in this study are time series measurements from a saltwater supply system located at Forskerparken in the port of Hirtshals. The variables recorded include the sea level, seawater temperature, pump performance (referred to as pump effect), and water intake level. Each variable was sampled at a frequency of \( 1 \: \mathrm{minute}^{-1} \) (one measurement per minute), providing high-resolution data for the system's operation. Table \ref{tab:data_description} provides summary statistics for the variables.

\input{describe_table}

This saltwater infrastructure plays a crucial role in supporting multiple stakeholders. It supplies saltwater internally to facilities within Forskerparken, including the North Sea Aquarium, while also servicing external commercial users such as Danish Salmon, the port of Hirtshals, Biomar, Vikingmar, and BinOcean. Effective management of the water intake level is essential to meet the demand from these downstream users and ensure consistent service levels. 

Among the variables in the dataset, the sea level in the harbor and the pump effect are hypothesized to be the primary factors influencing the water intake level. These variables are key to understanding and forecasting the intake dynamics. In addition, seawater temperature is included as a potential explanatory variable in the analysis. However, its contribution to the variability in water intake level is expected to be minimal and may serve only as a secondary factor.

Prior to conducting the analysis, the raw data were normalized to ensure consistency and comparability across variables. Specifically, all variables were scaled to the unit hypercube, such that each observation satisfies \( (\mathbf{x}, y) \in [0,1]^{d+1} \), where \( d \) represents the number of input features (in this case, sea level, temperature, and pump effect) and \( y \) corresponds to the water intake level. Unless otherwise specified, all data presented in this study reflect the normalized values. 

In the original dataset, there were some gaps in the recording of variables, due to occasional sensor downtime, and other technical issues.  Some of the measurement gaps were shorter than 60 minutes, and these gaps were imputed by linearly interpolating the missing values to maintain continuity in the time series. Larger gaps naturally divided the data into smaller dataset. Any dataset longer than 10 days were subdivided into segments of equal length, each between 5-10 days, to facilitate random assignment of datasets into training, validation, and test sets with a 60\%, 20\%, 20\% split. Although the time span from the first to the last measurement point was 582 days, gaps in the data reduced the total amount of usable measurements to 308 days. Of this, the 5-10 day datasets accounted for 243 days of data, with the remaining 63 days of data consisting of shorter datasets.

\subsection{Compression}
To establish a baseline, the dataset was first compressed using a lossless compression method to reduce storage requirements without compromising the integrity of the original data. Specifically, Brotli compression\footnote{\href{https://datatracker.ietf.org/doc/html/rfc7932}{https://datatracker.ietf.org/doc/html/rfc7932}} was employed. Brotli achieves high compression ratios by combining three core techniques: a variant of the LZ77 algorithm to identify and eliminate redundant patterns, Huffman coding to replace frequently occurring symbols with shorter codes, and second-order context modeling to improve the encoding of symbol probabilities based on the context. These techniques collectively enable efficient and effective lossless compression of the data.

For lossy compression, the Discrete Wavelet Transform (DWT) was applied using biorthogonal spline wavelets. This approach allows the decomposition of the time series data into different frequency components, enabling selective retention of the most significant features while discarding less relevant details. The use of biorthogonal spline wavelets is particularly advantageous due to their symmetry and smoothness properties, which ensure minimal distortion in the reconstructed data. Further details regarding the implementation and parameter selection of the DWT for lossy compression are provided in the subsequent sections.

\subsubsection{Discrete Wavelet Transform}
The Discrete Wavelet Transform (DWT) belongs to a broader class of Functional Approximation (FA) methods, which also include approaches such as Piecewise Polynomial Approximation (PPA), Discrete Cosine Transform (DCT), and Fourier Transform. These methods compress time series by approximating them with mathematical functions, offering interpretability and control over reconstruction error \cite{chiarot2023survey}. While each technique has its strengths, a detailed examination of their various aspects is beyond the scope of this paper.

The DWT decomposes a signal through a series of filtering operations followed by downsampling. At each resolution level, the signal is passed through complementary low-pass and high-pass filters, producing two sets of coefficients, namely \textit{approximation coefficients} ($a_j$), which represent the low-frequency components that capture the overall shape of the signal, and
\textit{detail coefficients} ($d_j$), which represent the high-frequency components that capture fine details and rapid changes.

For a signal at resolution level $j$, the decomposition process to the next resolution level $j+1$ can be mathematically expressed as:

\[a_{j+1}[n] = \sum_k h[k-2n] \: a_j[k], \,\,\,n=1,\dots, n_{j+1}\]
\[d_{j+1}[n] = \sum_k g[k-2n] \: a_j[k], \,\,\,n=1,\dots, n_{j+1}\]
\[n_{j+1}=n_j/2\]

where $a_{j+1}[n]$ are the approximation coefficients at level $j+1$, $d_{j+1}[n]$ are the detail coefficients at level $j+1$, $h[k]$ is the low-pass filter (scaling function), $g[k]$ is the high-pass filter (wavelet function) and the factor of 2 in the indexing ($k-2n$) represents the downsampling operation. This recursive process begins with the original signal as the initial approximation ($a_0[1], \dots, a_0[n_0]$) and continues to the desired decomposition level, creating a multi-resolution representation that efficiently captures both global trends and local features across different scales. 

For compression, after applying the DWT, we can threshold the detail coefficients, setting small values to zero:

\[d_j'[n] = \begin{cases}
d_j[n], & \text{if } |d_j[n]| > \epsilon \\
0, & \text{otherwise}
\end{cases}\]

This thresholding operation typically results in a sparse representation where only 10-20\% of the coefficients need to be stored, significantly reducing storage requirements while preserving the essential characteristics of the original time series upon reconstruction. Please refer to e.g. \cite{mallat_theory_1989,mallat_wavelet_2009} for more details on the discrete wavelet transform. 

In this study we use the \textit{biorthogonal} family of wavelets for compression. What makes biorthogonal wavelets particularly suitable for compression is that they satisfy the perfect reconstruction condition while allowing asymmetric filter designs. This means we can prioritize certain properties in the analysis (deconstruction) filters (like vanishing moments) while maintaining smoothness in the synthesis (reconstruction) filters \cite{https://doi.org/10.1002/cpa.3160450502}.

The biorthogonal wavelets are typically indexed by \texttt{BiorNr.Nd}, where \texttt{Nr} is the number of vanishing moments in the reconstruction wavelets, while \texttt{Nd} is the number of vanishing moments in the deconstruction wavelets. In this study we investigated the biorthogonal wavelets \texttt{Bior1.1}, \texttt{Bior1.5}, \texttt{Bior2.8}, \texttt{Bior3.9}, and \texttt{Bior6.8}.

%

\subsubsection{Compression Rate}
The lossy compression rate is calculated to quantify the extent of data reduction achieved after applying a threshold to the wavelet coefficients during the Discrete Wavelet Transform (DWT). Specifically, the lossy compression rate, denoted as \( r_{\mathrm{lossy}} \), is defined as the proportion of wavelet coefficients that are zero (\( N_{0} \)) relative to the total number of samples in the uncompressed time series (\( N_Y \)). Mathematically, this can be expressed as:

\[
r_{\mathrm{lossy}} = \frac{N_{0}}{N_Y}
\]

\subsubsection{Normalized Mutual Information} \label{sec:nmi}
To evaluate the quality of the reconstructed, compressed dataset \( \tilde{Y} \), we measure the remaining signal strength using Normalized Mutual Information (NMI). The NMI quantifies the amount of shared information between the compressed and uncompressed datasets, normalized to ensure interpretability. It is defined as:

\[
\mathrm{NMI}(r) = \frac{I(\tilde{Y}_{r}; Y)}{I(\tilde{Y}_{0}; Y)}
\]

where \( r \) represents the compression rate, \( Y \) is the (distribution of) the target variable in the uncompressed dataset and \( I(X; Y) \) denotes the mutual information between two random variables \( X \) and \( Y \).

The mutual information \( I(X; Y) \) measures the reduction in uncertainty about \( Y \) given knowledge of \( X \). It is formally defined as:

\[
I(X; Y) = D_{\mathrm{KL}}(P_{(X, Y)} \parallel P_X \otimes P_Y)
\]

where \( D_{\mathrm{KL}} \) is the Kullback-Leibler (KL) divergence, which quantifies the difference between two probability distributions and \( P_{(X, Y)} \) is the joint distribution of \( X \) and \( Y \), while \( P_X \otimes P_Y \) is the product of their marginal distributions.

For the NMI, the numerator \( I(\tilde{Y}_{r}; Y) \) represents the mutual information between the compressed signal \( \tilde{Y}_{r} \) (at compression rate \( r \)) and the uncompressed signal \( Y \). The denominator \( I(\tilde{Y}_{0}; Y) \) is the mutual information between the original, uncompressed signal \( \tilde{Y}_{0} \) and itself, effectively acting as a normalization factor.

This normalization ensures that the NMI satisfies the following properties:
\begin{itemize}
\item \( \mathrm{NMI}(r) \in [0, 1] \): The NMI is bounded between 0 and 1.
\item \( \mathrm{NMI}(0) = 1 \): When there is no compression (\( r = 0 \)), the compressed signal perfectly retains all information from the original signal.
\item \( \mathrm{NMI}(1) = 0 \): When the compression rate is maximal (\( r = 1 \)), the compressed signal retains no information from the original signal.
\end{itemize}
By using NMI, we are able to cast the remaining signal strength in a way that directly relates the compressed and uncompressed datasets, providing an interpretable, dimensionless metric to assess the impact of compression.

\subsection{Time series modeling}
In this study, we employed three distinct methods for time series forecasting: (1) an Ordinary Least Squares (OLS) regression approach (i.e., multiple linear regression) \cite[see, e.g.,][]{keele_dynamic_2006}, (2) an XGBoost regressor \cite{chen_xgboost_2016}, and (3) a Time Series Dense Encoder (TiDE) regressor \cite{das_long-term_2024}. These methods were selected for their ability to model the temporal dependencies inherent in time series data. All three approaches can be broadly categorized as Lagged Dependent Variable (LDV) models, as they incorporate lagged observations of the dependent variable as predictors. The general framework and principles of LDV models are discussed in greater detail in the following section.

In preliminary experiments, we also evaluated the performance of several other deep learning-based methods for time series forecasting. The methods tested included DeepAR \cite{SALINAS20201181}, the Temporal Fusion Transformer (TFT) \cite{LIM20211748}, and N-HiTS \cite{challu_nhits_2023}. Despite their widespread use and effectiveness in certain predictive tasks, these methods consistently underperformed when compared to TiDE in the context of this study. Consequently, they were excluded from further analysis to maintain a focused comparison of the most effective models.

We used the free and open source Python library \texttt{Darts} \cite{JMLR:v23:21-1177} for the numerical experiments.

\subsubsection{Lagged Dependent Variable Model}\label{sec:ldv}
The Lagged Dependent Variable (LDV) approach is a widely used framework in time series modelling that incorporates past values of the dependent variable as predictors to capture temporal dynamics. This method is particularly effective for forecasting tasks where the goal is to predict future values of a time series while accounting for historical patterns and potential influences of covariates. Below, we formalize the LDV approach, extend it to a multi-horizon forecasting setup, and include both past and future known covariates.

Let \( y_t \) denote the dependent variable at time \( t \), and let \( \mathbf{x}_t \) represent a vector of covariates observed at time \( t \). The LDV model can be expressed as:

\[
y_t = f_{\boldsymbol{\theta}}(y_{t-1}, y_{t-2}, \ldots, y_{t-p}, \mathbf{x}_{t}, \mathbf{x}_{t-1}, \ldots, \mathbf{x}_{t-q}) + \varepsilon_t,
\]

where \( f_{\boldsymbol{\theta}}(\cdot) \) is a potentially nonlinear function parameterized by \(\boldsymbol{\theta}\), \( p \) is the number of lags for the dependent variable included in the model, \( q \) is the number of lags for the covariates considered, and \( \varepsilon_t \) is the error term, typically assumed to follow a Gaussian distribution. The inclusion of lagged terms \( y_{t-1}, y_{t-2}, \ldots \) allows the model to capture autocorrelation in the series, while the incorporation of covariates \( \mathbf{x}_{t}, \mathbf{x}_{t-1}, \ldots \) facilitates the modeling of external effects on \( y_t \).

In multi-horizon forecasting, the objective is to predict the future values \( \{y_{t+1}, y_{t+2}, \ldots, y_{t+H}\} \) for a forecast horizon \( H \). Using the LDV approach, this can be achieved either iteratively (or ``autoregressively``) or directly. In the autoregressive approach, the model is first used to predict \( y_{t+1} \), which is then treated as an input to predict \( y_{t+2} \), and so on up to \( y_{t+H} \). This process relies on the recursive usage of lagged predictions:
   \[
   \hat{y}_{t+h} = f_{\boldsymbol{\theta}}(\hat{y}_{t+h-1}, \hat{y}_{t+h-2}, \ldots, y_{t+h-p}, \mathbf{x}_{t+h}, \mathbf{x}_{t+h-1}, \ldots, \mathbf{x}_{t+h-q}),
   \]
   for \( h = 1, 2, \ldots, H \).

Conversely, for the direct approach, separate models are trained for each forecast horizon \( h \), where the dependent variable of interest is \( y_{t+h} \). The formulation for each horizon is:
   \[
   y_{t+h} = f_{\boldsymbol{\theta_h}}(y_t, y_{t-1}, \ldots, \mathbf{x}_{t+h}, \mathbf{x}_{t+h-1}, \ldots) + \varepsilon_{t+h}.
   \]

In many practical time series applications, certain covariates are known in advance for future time steps. Examples include calendar effects, policy and control interventions, or exogenous variables such as weather forecasts. The LDV model can integrate these future-known covariates \( \mathbf{x}_{t+h} \) into the predictive framework. Specifically, the function \( f_{\boldsymbol{\theta}}(\cdot) \) can be designed to utilize both past covariates \( \mathbf{x}_{t}, \mathbf{x}_{t-1}, \ldots \) and future covariates \( \mathbf{x}_{t+h} \) to enhance forecasting accuracy when such future known covariates are available.

Depending on the complexity of the time series, \( f_{\boldsymbol{\theta}}(\cdot) \) can be linear (e.g., an ordinary least squares (OLS) regression model) or nonlinear (e.g., a neural network or tree-based model). In this study, we use both a linear regression model \cite[see e.g.][]{keele_dynamic_2006} and an XGBoost regressor \cite{chen_xgboost_2016} as $f_{\boldsymbol{\theta}}(\cdot)$. Note that the deep learning model we employ, the Time Series Dense Encoder (TiDE), can also broadly be considered as an LDV approach. However, this neural network approach involves some particular techniques unique to the method and is therefore described in more detail below.

\subsubsection{TiDE}
The Time Series Dense Encoder \cite[TiDE][]{das_long-term_2024} is a neural network architecture specifically designed for time series forecasting, prioritizing computational efficiency, scalability, and robust predictive performance. TiDE implements a fully connected feedforward architecture, avoiding the use of recurrent, convolutional, or transformer architectures, which are common in other neural network based time series models. This simple design choice results in significantly faster training and inference times than other architectures, making the approach well-suited for large-scale and long-term forecasting tasks. The architecture utilizes a multi-layer perceptron (MLP)-based encoder to process historical time series data, extracting dense representations of temporal patterns from fixed-length input windows. These representations are then passed to dedicated decoder modules, which are responsible for generating forecasts over specified horizons. To enhance generalization and stability during training, the model incorporates various architectural improvements such as residual connections \cite{7780459} and dropout regularization \cite{JMLR:v15:srivastava14a}.

\section{Results}
\subsection{Fitting the models to the uncompressed data} \label{sec:results_uncompressed}
We evaluated the performance of three models -- Ordinary Least Squares (OLS), XGBoost, and TiDE -- by fitting them to the uncompressed dataset. The target variable for forecasting was the water intake level. As past input covariates, we included the sea level and water temperature, which were only known up to the beginning of the forecasting period. Additionally, the pump effect, a future-known covariate, was provided as input to the models during the forecasting process. Note that the sea level can be predicted using e.g. tide tables, and some approximation could thus have been included as a future covariate to additionally improve the forecasting models presented here. However, in this study it was included only as a past covariate for simplicity.

Each model was trained to forecast the water intake level for a one-hour horizon using information from the previous six hours. Specifically, the inputs consisted of six hours of historical data for both the target variable and all the covariates. Given a sampling frequency of \( 1 \: \mathrm{minute}^{-1} \), a single time series datapoint was composed of \( n_{\mathrm{data}} = 420 \) time steps, representing \( 7 \times 60 \) minutes (seven total hours of data).

\begin{figure}[htbp]
    \centering
    \includegraphics[width=0.8\linewidth]{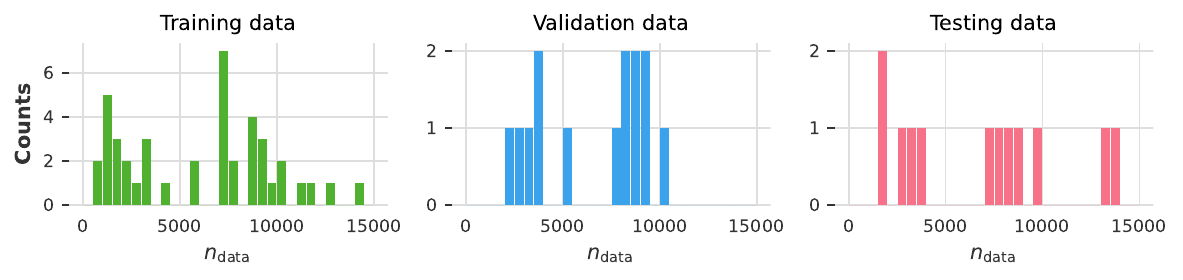}
    \caption{Dataset sizes for the training, validation and testing datasets.}
    \label{fig:dataset_sizes}
\end{figure}

To train the models, we utilized \( N_{\mathrm{train}} = 42 \) time series datasets and validated their performance using \( N_{\mathrm{validation}} = 14 \) datasets. Both training and validation datasets contained sufficient time steps to extract at least one complete time series per dataset. The sizes of the datasets varied due to differences in the lengths of recorded time series. Figure \ref{fig:dataset_sizes} illustrates the distribution of dataset sizes across the training, validation, and testing datasets, providing an overview of the data used in the modeling process.

The model evaluation was extended to a six-hour forecast horizon, corresponding to \( n_{\mathrm{forecast}} = 360 \) time steps, consistent with the sampling frequency of \( 1 \: \mathrm{minute}^{-1} \). Since the models were trained to directly predict only a one-hour horizon, the remaining five hours of the forecast were generated using an autoregressive approach. This method involves iteratively feeding the model’s own predictions for the current time step back into the input to forecast subsequent time steps, as described in Section \ref{sec:ldv}.

Figure \ref{fig:rmse_all} presents the distribution of Root Mean Squared Errors (RMSEs) for each model across the \( N_{\mathrm{test}} = 12 \) testing datasets. The results for the uncompressed scenario is presented as a compression rate of \( r = 0.0 \). For additional context, the first row of Table \ref{tab:results} provides the overall baseline metrics -- RMSE and mean absolute error (MAE) -- calculated as averages over all testing data. The results indicate that the three models exhibit broadly comparable performance on the baseline uncompressed data. Notably, the OLS model achieves the lowest overall error metrics, suggesting slightly superior predictive accuracy in this case.

\begin{figure}[htbp]
    \centering
    \includegraphics[width=0.9\linewidth]{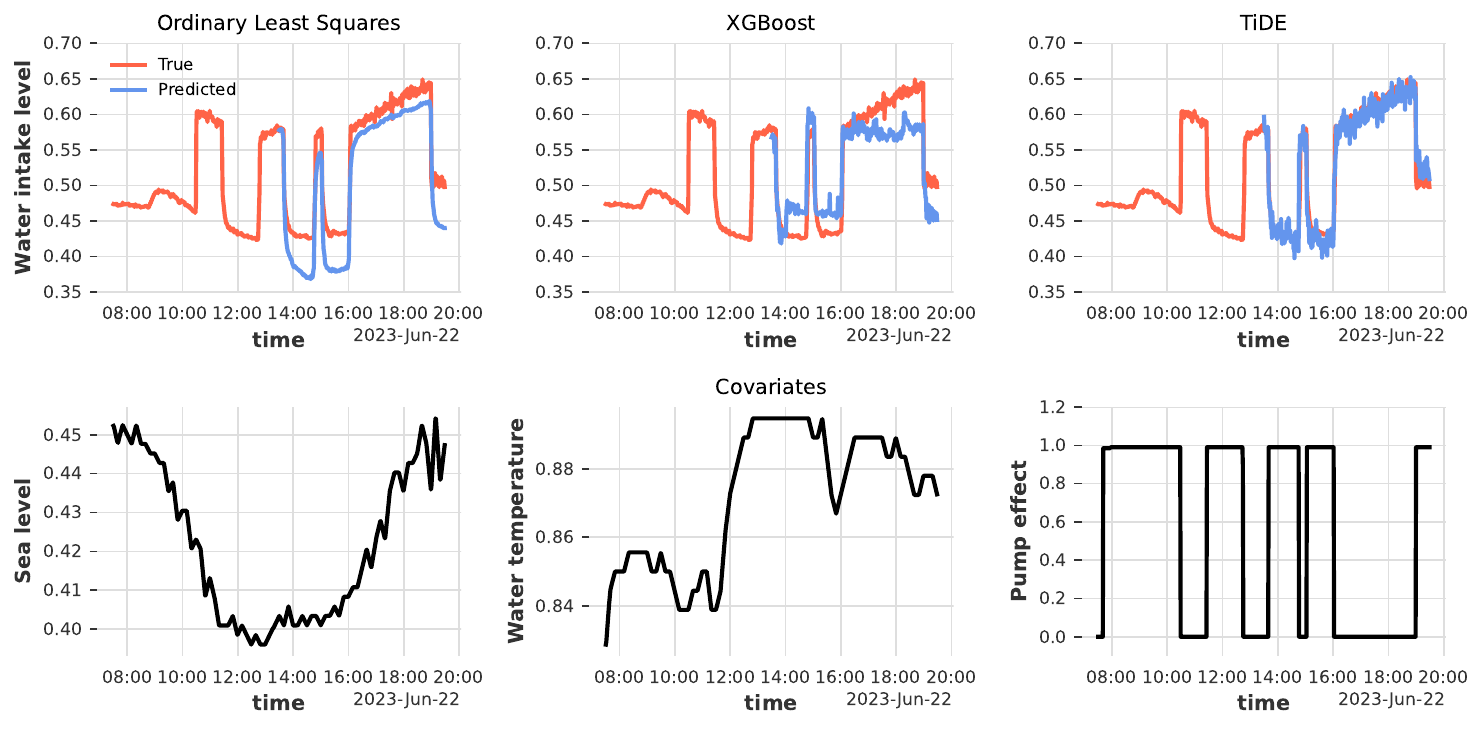}
    \caption{Example of model predictions for all 3 time series forecasting models. Each model used a context of six hours to make predictions for another six hours, of which one hour was directly predicted and the remaining was computed autoregressively.}
    \label{fig:prediction_example}
\end{figure}

To further illustrate model performance, Figure \ref{fig:prediction_example} showcases examples of forecasted time series from the testing dataset. These examples show the models’ ability to capture key trends and fluctuations in the water intake level, providing a visual comparison of their predictive capabilities.


\subsection{Fitting the models to the compressed data}\label{sec:results_compressed}
For the lossless compression of our dataset, we utilized the Brotli algorithm, achieving a compression rate of \( r_{\mathrm{lossless}} = 0.36 \). As a result, lossy compression rates satisfying \( r_{\mathrm{lossy}} < 0.36 \) are not of practical interest in this study, as they would fail to outperform the efficiency of the lossless compression baseline.

\begin{figure}[htbp]
    \centering
    \includegraphics[width=\linewidth]{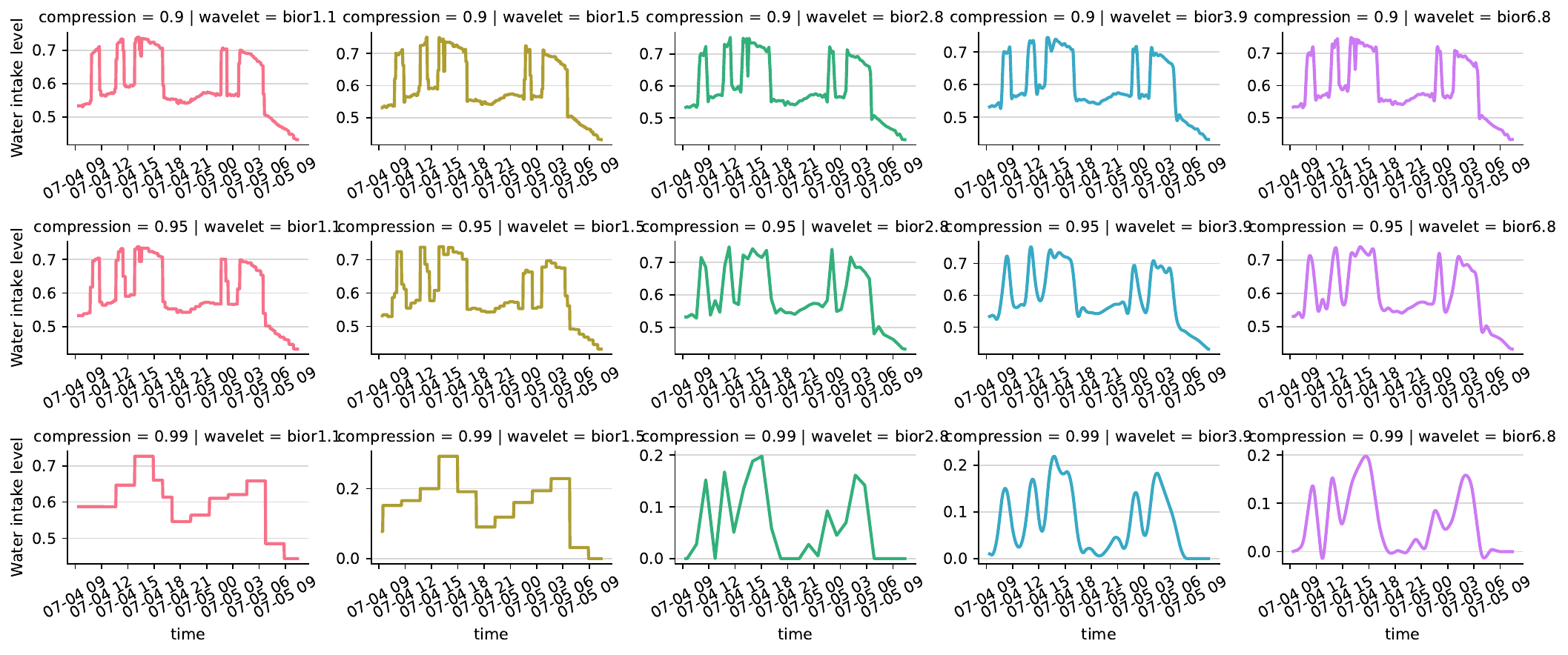}
    \caption{Example subset of the compressed, normalized target variable. The rows show different compression rates $r \in \{0.9, 0.95, 0.99\}$ while the columns show different families of biorthogonal wavelets.}
    \label{fig:compression_example}
\end{figure}

In our analysis of lossy compression, we employed the discrete wavelet transform (DWT) and investigated a range of compression rates, specifically \( r_{\mathrm{lossy}} \in \{0.4, 0.6, 0.8, 0.9, 0.95, 0.99, 0.999\} \). These rates were selected to assess the trade-off between data fidelity and compression efficiency across varying levels of compression. Figure \ref{fig:compression_example} illustrates an example from the dataset, showing the normalized, compressed target variable (water intake level) for a subset of compression rates: \( r_{\mathrm{lossy}} \in \{0.9, 0.95, 0.99\} \). 

From the examples in Figure \ref{fig:compression_example}, it is evident that as the compression rate increases, distinct qualitative artifacts emerge in the reconstructed time series due to the characteristics of the wavelets. Broadly, wavelets with rougher profiles, such as \texttt{Bior1.1} and \texttt{Bior1.5}, demonstrate a better ability to preserve abrupt changes (shocks) in the original time series. However, this comes at the cost of erasing medium-frequency oscillations, leading to a loss of detail. In contrast, smoother wavelets, such as \texttt{Bior3.9} and \texttt{Bior6.8}, retain more of the medium-frequency oscillations present in the original data. This preservation, however, comes with trade-offs: the smoothing effect reduces the prominence of shocks, and in some cases, introduces Gibbs artifacts \cite[see e.g.][]{d65b784a09324920977631d10fd7b86d}, which manifest as spurious oscillations near discontinuities.

This behavior illustrates trade-offs associated with wavelet selection in lossy compression. Rougher wavelets are more suited for applications prioritizing the retention of sharp transitions, while smoother wavelets may be preferable in scenarios where preserving medium-length oscillations is important, even if some artifacts are introduced.

For the analysis involving compressed data, we used the same experimental setup described in Section \ref{sec:results_uncompressed}. Specifically, each model was trained and validated using datasets subjected to lossy compression and tested on the corresponding uncompressed data. This setup allowed us to evaluate the models' ability to perform accurate forecasting when trained on compressed data, reflecting scenarios where storage or transmission constraints necessitate compression.

\begin{figure}[htbp]
    \centering
    \includegraphics[width=\linewidth]{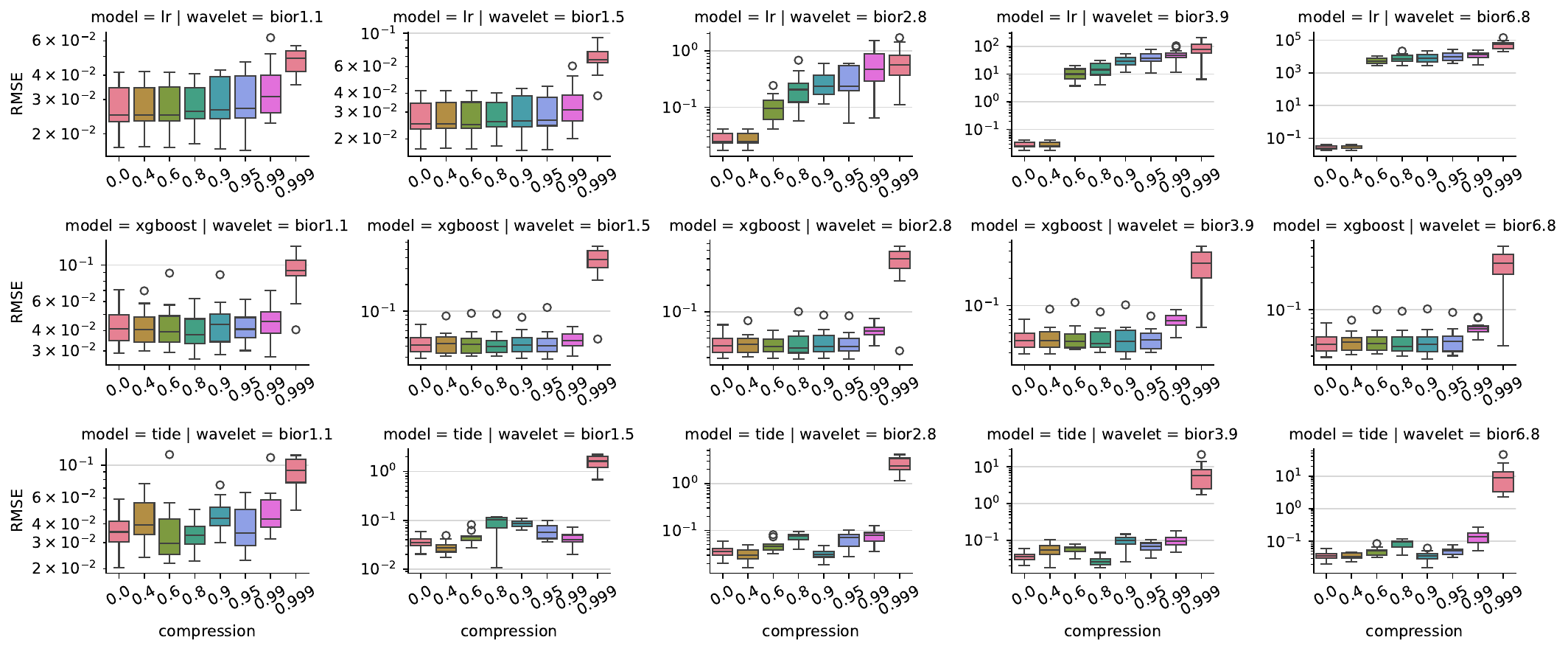}
    \caption{RMSE for every model against the compression rate $r$. The rows show different models, OLS, XGBoost and TiDE, while the columns show different families of biorthogonal wavelets. The boxplots show the distribution of RMSE's across the $N_{\mathrm{test}}=12$ different testing datasets, including outliers. The whiskers represent the 1.5 Interquantile Range (IQR).}
    \label{fig:rmse_all}
\end{figure}

Figure \ref{fig:rmse_all} shows the distribution of Root Mean Squared Errors (RMSEs) across all combinations of models, wavelet types, and compression rates. These results are computed using the \( N_{\mathrm{test}} = 12 \) testing datasets. The figure provides a comparison of how the performance of each model changes with the choice of wavelet and compression rate. In addition, Table \ref{tab:results} lists the overall error metrics -- RMSE and mean absolute error (MAE) -- for each combination of model, wavelet, and compression rate, averaged across all testing datasets.

The results show that model performance is influenced by the wavelet type and compression rate. As compression rates increase (meaning more information is removed), errors generally increase, reflecting the trade-off between compression and data fidelity. Different wavelets also affect performance in distinct ways. For example, as explained in more detail above, rougher wavelets tend to remove fine details but preserve sharp transitions, while smoother wavelets retain more detail but can introduce artifacts. These factors affect the prediction accuracy in various ways and must be carefully considered when applying lossy compression to datasets used for forecasting.

For the OLS model, performance remains strong when applied to uncompressed data and data compressed using the rougher wavelets (\texttt{Bior1.1} and \texttt{Bior1.5}). However, when smoother wavelets (\texttt{Bior3.9} and \texttt{Bior6.8}) are used, and the compression rate is increased, both the MAE and RMSE for the OLS model increase significantly. When the OLS model is trained on smoothed data, it relies on short-lag inputs as predictors for future timesteps, as these inputs appear highly correlated with future values in the training data. However, this reliance leads to a model that is sensitive to small perturbations in the short-lag inputs. When the model encounters testing data containing noise or high-frequency oscillations in the short-lag inputs, these perturbations propagate through the model, resulting in large errors in the predictions.

One potential way to address the sensitivity of the OLS model to perturbation would be to also apply smoothing to the testing data, aligning its characteristics with the training data. This approach could reduce the impact of noise and high-frequency oscillations, making the predictions more stable. However, smoothing the testing data would also attenuate sharp transitions (shocks) in the signal, which are an important part of the underlying dynamics being modeled. Another possible solution would involve introducing regularization techniques, such as ridge regression, to stabilize the OLS model by penalizing large coefficients. However, exploring these alternatives is beyond the scope of the present study.

The XGBoost model, while not achieving the same level of performance as the best OLS model on uncompressed data, demonstrates greater robustness to compression artifacts the introduced by smoother wavelets. Its performance remains stable across most wavelets and compression rates, showing notable degradation only at extreme compression rates (\( r > 0.99 \)). Even under these conditions, the deterioration in performance is less severe compared to the OLS model. The consistency of the XGBoost model across different wavelets and compression rates indicates that it is well-suited for scenarios where data compression is necessary, particularly when either smooth wavelets or extreme compression rates are used.

The performance of the TiDE model does not exhibit a clear pattern across wavelets and compression ratios. While it is generally more robust to compression artifacts than the OLS model, its performance varies depending on the specific wavelet and compression rate. This variability can likely be attributed to the inherent sources of randomness in deep learning models \cite[see, e.g.,][]{Bengio2012, Goodfellow-et-al-2016, beam_challenges_2020}. Factors such as random initialization of weights, stochastic optimization processes, and differences in training dynamics can lead to variations in outcomes, even when using the same data. Despite this variability, the TiDE model delivers competitive performance across all examples, except at $r = 0.999$.


\input{results_table}

\subsection{Normalized Mutual Information}
\begin{figure}[htbp]
    \centering
    \includegraphics[width=0.7\linewidth]{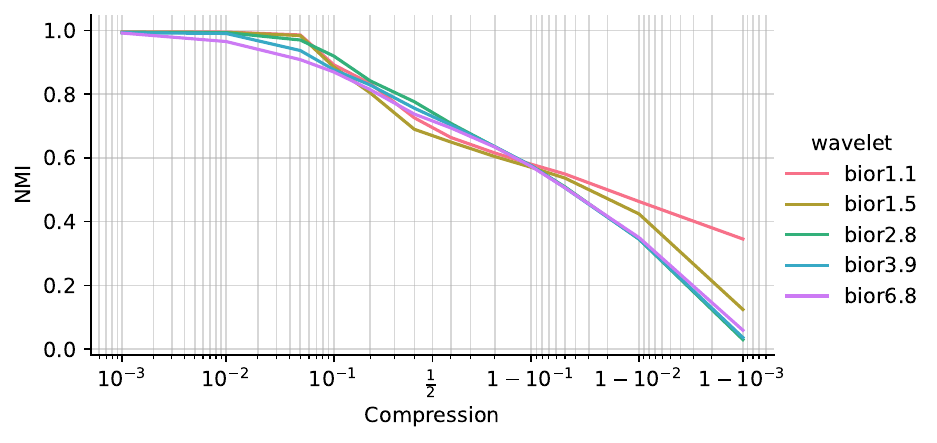}
    \caption{Normalized Mutual Information (NMI) for each wavelet as the compression is increased. Note that the x-axis is displayed on a \textit{logit}-scale.}
    \label{fig:nmi_by_compression}
\end{figure}

The mutual information was computed using entropy estimation based on a \( k \)-nearest neighbors approach \cite{1987PrIT...23...95K, PhysRevE.69.066138, 10.1371/journal.pone.0087357}, with \( k = 10 \) neighbors. This method was implemented using the \texttt{scikit-learn} library \cite{scikit-learn}. Figure \ref{fig:nmi_by_compression} displays the Normalized Mutual Information (NMI; see Section \ref{sec:nmi}) for each wavelet as a function of the compression rate.

At very low compression rates (\( r \) close to 0, which are not of practical interest in this study), the coarser wavelets (\texttt{Bior1.1} and \texttt{Bior1.5}) retain the highest amount of mutual information from the original data. As the compression rate increases to medium values (\( 0.2 < r < 0.8 \)), smoother wavelets (\texttt{Bior2.8}, \texttt{Bior3.9}, and \texttt{Bior6.8}) retain more information compared to the coarser wavelets. Finally, at higher compression rates (\( r > 0.8 \)), the rougher wavelets (\texttt{Bior1.1} and \texttt{Bior1.5}) again retain more information than the smoother wavelets.

We modeled the Normalized Mutual Information (NMI) as a function of the compression rate \( r \) using regularized, incomplete beta functions \cite{NIST:DLMF}. Specifically, the NMI was expressed as \( \mathrm{NMI}(r) = 1 - I_r(\alpha, \beta) \), where \( I_r(\alpha, \beta) \) denotes the regularized incomplete beta function. To identify the characteristic shape of the NMI curve for each wavelet, we performed a least squares fit to determine the parameters \( (\alpha, \beta) \).

The parameter \( \alpha \) primarily governs the steepness of the NMI curve as \( r \to 0^+ \), while \( \beta \) controls the steepness as \( r \to 1^- \). Lower values of \( \alpha \) and \( \beta \) correspond to steeper declines in the NMI near the respective limits. Ideally, a desirable compression scheme would have a high \( \alpha \) and a low \( \beta \), indicating that most information is retained as the compression rate increases, with a sharp drop-off only as \( r \to 1^- \).

\begin{figure}[htbp]
    \centering
    \includegraphics[width=\linewidth]{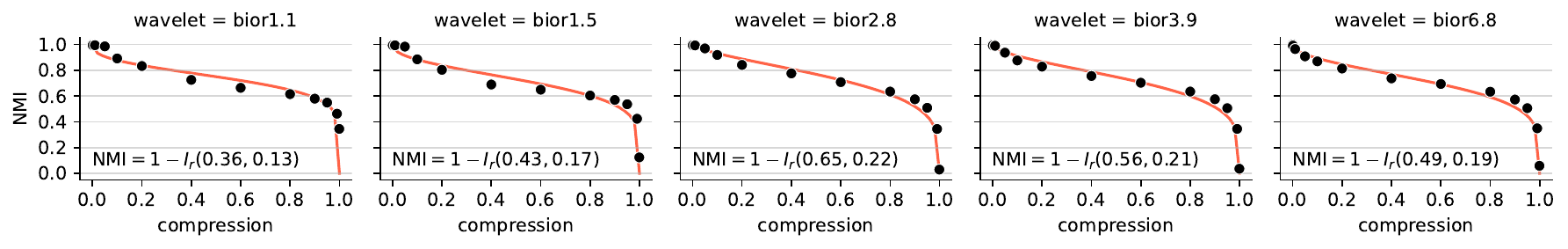}
    \caption{Normalized Mutual Information (NMI) for each wavelet as the compression is increased, with regularized, incomplete beta functions fitted to the data from each wavelet. Each panel also shows the function of the fitted curve.}
    \label{fig:nmi_by_wavelet}
\end{figure}

Figure \ref{fig:nmi_by_wavelet} illustrates the fitted NMI curves for each wavelet. Across the five wavelets, no single wavelet clearly outperforms the others. The \texttt{Bior1.1} wavelet exhibits the highest steepness at both \( r \to 0^+ \) and \( r \to 1^- \), indicating it erases the most information at low compression rates, but also preserves the most information at high compression rates. Conversely, \texttt{Bior2.8} displays the lowest steepness at both \( r \to 0^+ \) and \( r \to 1^- \), suggesting it retains the most information at low and medium compression rates but loses more information as the compression rate approaches 1. This indicates that \texttt{Bior2.8} is better suited for retaining information at moderate compression, while \texttt{Bior1.1} is more effective at preserving information at very high compression rates.

\begin{figure}[htbp]
    \centering
    \includegraphics[width=\linewidth]{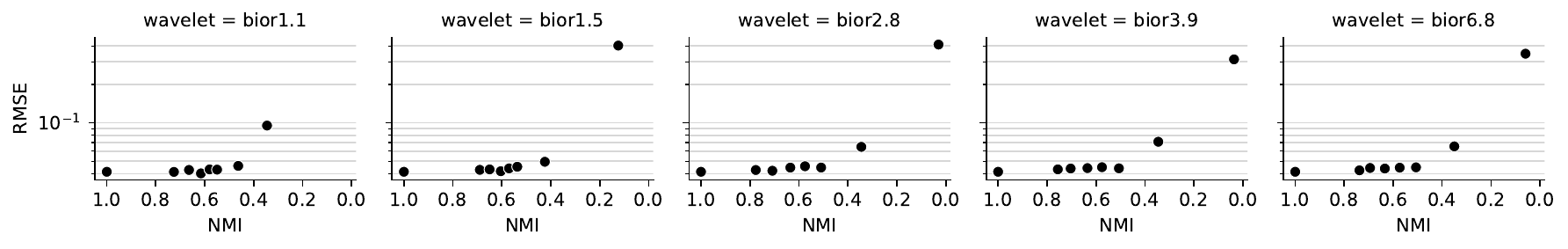}
    \caption{RMSE versus NMI for the XGBoost model. Note that the x-axis is inverted.}
    \label{fig:rmse_by_nmi}
\end{figure}

When analyzing the most robust time series regression model, XGBoost (see Section \ref{sec:results_compressed}), we observe a clear correspondence between the Normalized Mutual Information (NMI) and the Root Mean Squared Error (RMSE), as shown in Figure \ref{fig:rmse_by_nmi}. This relationship is nonlinear: the RMSE remains relatively stable as long as the NMI is high but increases sharply when the NMI drops below a critical threshold. This sudden rise in RMSE aligns with the steep decline in NMI at high compression rates, corresponding to the rightmost notch in the NMI versus compression rate \( r \) described in the previous section.

This observation suggests that the choice of compression rate has a direct impact on predictive performance, which can be somewhat anticipated by analysing the NMI. To balance information retention and compression efficiency, we recommend using an "elbow" method with the RMSE versus compression rate graph \cite[see, e.g.,][]{shi_quantitative_2021} to select the compression rate. This method identifies the point at which further increases in compression lead to diminishing returns in terms of information retention, helping to avoid the sharp deterioration in model performance associated with excessive compression.

\section{Discussion}
This paper examines the impact of lossy data compression techniques, particularly the Discrete Wavelet Transform (DWT), to time series forecasting for smart energy systems. In this context, the relationships between the wavelet order, compression rate, forecasting model, and forecasting performance were investigated.

Data compression is essential for reducing the storage and transmission requirements, especially in systems with moderate to high-frequency data collection, such as the seawater supply system analyzed in this study. Lossless compression methods, such as the Brotli algorithm, preserve all original information but achieve only moderate compression rates. In contrast, lossy compression using the DWT can achieve much higher compression rates by discarding some details from the signal. However, this reduction in data size comes at the cost of introducing artifacts in the reconstructed data and, consequently, affect the performance of forecasting models. Balancing storage efficiency with data fidelity is therefore a key challenge when applying lossy compression.

The choice of wavelet order significantly impacts the quality of the compression. Rougher wavelets, such as \texttt{Bior1.1}, are better suited to retaining sharp transitions in the data, making them effective for signals with sudden changes. Smoother wavelets, such as \texttt{Bior6.8}, preserve medium-frequency oscillations but smooth out sharp transitions and can introduce oscillatory artifacts near discontinuities, also known as Gibbs phenomena. The effectiveness of a wavelet depends on the specific characteristics of the dataset. For example, rougher wavelets are useful when sharp transitions are important, whereas smoother wavelets are more effective for capturing smoother patterns or medium-frequency oscillations. Selecting a wavelet that aligns with the properties of the time series is critical for achieving effective compression.

The performance of forecasting models varies depending on the level of compression and the order of the wavelet. The Ordinary Least Squares (OLS) model performed well on uncompressed data but was highly sensitive to distortions introduced by compression, particularly when smoother wavelets were used. This sensitivity arises from the reliance of the OLS model on short-lag inputs, which are more affected by the artifacts introduced during compression. Regularization techniques or preprocessing of the testing data could potentially alleviate this issue, but these approaches were beyond the scope of this study. The XGBoost model demonstrated greater robustness to compression artifacts and maintained stable performance across a wide range of compression rates and wavelet types. This consistency makes it a suitable choice for applications requiring compressed data. The Time Series Dense Encoder (TiDE) model showed competitive performance overall but exhibited variability in its results, which may be attributed to the inherent randomness in deep learning models. Despite this variability, TiDE remained a viable option for forecasting under compressed conditions.

Normalized Mutual Information (NMI) provided a quantitative measure of the effects of compression on the data fidelity. NMI decreased nonlinearly with increasing compression rates, with a sharp drop observed at higher rates. This decline in NMI correlated with a significant increase in prediction errors, especially when the retained information fell below a critical threshold. The results suggest that NMI is a useful metric for guiding the selection of compression rates. By analyzing the NMI-compression curve, an optimal compression rate can be identified that balances information retention and storage efficiency. This approach can assist in avoiding the steep loss of predictive accuracy associated with excessive compression.

The findings have several practical implications for the operation of smart energy systems. High-frequency data collection in such systems generates large volumes of data that need to be stored or transmitted efficiently. Compression using DWT provides a way to reduce storage or transmission requirements while retaining the data features necessary for forecasting. The study also demonstrates that predictive performance depends on aligning wavelet properties with the structure of the data, and that forecasting performance can be maintained even with compressed data, provided that the compression is carefully configured and the appropriate forecasting model is used.

A key limitation of the proposed compression approach is the potential loss of information that was not initially considered critical for the intended forecasting task. Once compression is applied, features or patterns that could prove valuable for other downstream applications—such as anomaly detection, system diagnostics, or alternative predictive models—may be irreversibly discarded. This is relevant for lossy compression methods like the Discrete Wavelet Transform (DWT), where decisions about retaining or discarding data are task-specific and often guided by pre-defined assumptions about which features are most relevant.

This limitation underscores the importance of adopting a holistic approach when designing compression strategies, ensuring that decisions about data retention account for the broader ecosystem of potential downstream applications. Without such foresight, compression risks narrowing the utility of the data, limiting its adaptability to unforeseen analytical needs. Balancing immediate efficiency with future flexibility is therefore a critical challenge in the design of data-driven systems.

\subsection{Implementation}
While this study addresses the feasibility of lossy data compression for time series forecasting, it is equally important to consider how such compression can be implemented in practical end-to-end systems. In real-world deployments, such as IoT-based smart energy systems, the end-to-end pipeline typically involves sensors capturing high-frequency data, transmitting it over a network, and storing it in centralized or distributed repositories for downstream processing and analysis. Incorporating compression into this pipeline introduces opportunities for significant efficiency gains, particularly when implemented at the edge (i.e., on the sensor itself).

Performing data compression directly on the sensor hardware offers several advantages. First, it reduces the volume of data that needs to be transmitted over potentially bandwidth-constrained networks, such as low-power wide-area networks (LPWANs) or cellular IoT networks. This reduction can significantly lower transmission costs and improve the reliability of data transfer in scenarios where network latency or packet loss is a concern. Second, reducing the data size at the source alleviates the need for extensive storage capacity on the sensor or intermediate gateways, making the overall system more lightweight and scalable.

To enable compression on the sensor itself, lightweight and computationally efficient algorithms such as the Discrete Wavelet Transform (DWT) can be embedded into the sensor firmware. Modern microcontrollers and edge computing platforms used in IoT devices, such as ESP32 or ARM Cortex-M series processors, are typically capable of handling simple wavelet transforms in real-time due to their optimized signal processing capabilities. Furthermore, many sensors already incorporate basic preprocessing capabilities, such as filtering and data smoothing, which could be extended to include compression routines.

While performing compression at the sensor level is advantageous, it also introduces some challenges. Sensors operating in resource-constrained environments often have limited computational power, memory, and energy budgets. Therefore, selecting an appropriate wavelet family and compression rate is critical to ensure that the computational overhead remains minimal while maintaining sufficient data fidelity for downstream forecasting tasks.

Another opportunity lies in adaptive compression, where the compression rate or wavelet type is dynamically adjusted based on the characteristics of the incoming data. For instance, during periods of low variability in the signal, higher compression rates could be used, whereas during periods of rapid change, lower compression rates might be preferable to retain critical details. Developing lightweight algorithms for such adaptive compression would enhance the flexibility and utility of sensor-level data processing.

Implementing compression on the sensor itself aligns with the broader trend of edge computing, where data processing is moved closer to the source of data generation. This paradigm not only reduces the reliance on centralized infrastructure but also enhances data privacy by minimizing the transfer of unprocessed raw data. Additionally, edge-based compression can extend the battery life of sensors in remote or hard-to-reach locations by reducing energy consumption associated with data transmission.

Overall, the integration of wavelet-based compression techniques directly into sensor hardware represents an interesting avenue of research for enabling scalable and efficient time series forecasting systems. By reducing the resource demands of data storage and transmission, this approach can facilitate the deployment of IoT solutions across a wide range of applications, from smart grids and water management systems to industrial automation and environmental monitoring.

\section{Conclusion and Future Work}
This study demonstrates that wavelet-based compression can support efficient time series forecasting in smart energy systems. By carefully balancing compression rates, data fidelity, and model performance, it is possible to achieve significant storage savings without compromising the accuracy of forecasts. This approach can improve the scalability and efficiency of data-driven decision making in systems where data storage and transmission constraints are a critical factor.

Future work could explore adaptive wavelet selection, which dynamically adapts to the characteristics of incoming data, to improve compression performance. The scope could also be extended to compare the DWT to other lossy compression methods such as other FA techniques, or simply reducing the polling frequency of the sensor. Hybrid methods that combine wavelet-based approaches with other compression techniques, such as dictionary-based or deep learning-based methods, could also be investigated. Extending the analysis to additional domains, such as climate modeling or financial forecasting, could help generalize these findings and identify domain-specific considerations.

\section{Acknowledgments}
This study was funded through EFFORT, Elforsk project ELF221-496872. We used Large Language Models (LLMs) as a writing aid for this paper.

\printbibliography


\end{document}

%% file: describe_table.tex
\begin{table}[htbp]
    \centering
    \caption{Summary statistics of the variables in the dataset, $n_{\mathrm{data}} = 1753804$}
    \label{tab:data_description}
    \begin{tabular}{lrrrr}
        \toprule
         & \textbf{Water Intake Level (cm)} & \textbf{Pump Effect (\%)} & \textbf{Sea Level (cm)} & \textbf{Water Temperature ($^{\circ}$C)} \\
        \midrule
        Arithmetic Mean & 192.6 & 84.2 & 1.0 & 7.0 \\
        Standard Deviation & 68.0 & 31.1 & 36.9 & 4.0 \\
        \midrule
        Min & 63.8 & 0.0 & -169.0 & 1.5 \\
        25\% & 134.0 & 86.0 & -17.1 & 4.8 \\
        50\% & 202.2 & 99.5 & 4.4 & 5.8 \\
        75\% & 236.4 & 99.5& 22.8 & 7.8 \\
        Max & 407.8 & 100.0 & 237.6 & 19.4 \\
        \bottomrule
    \end{tabular}
\end{table}

%% file: results_table.tex
\begin{table}[htbp]
\centering
\caption{RMSE and MAE for all combinations of models and compressions, including the baseline case.}
\label{tab:results}

\begin{tabular}{lrrrrrrr}
\toprule
\textbf{Wavelet} & $r_{\mathrm{lossy}}$ & \multicolumn{2}{r}{\textbf{OLS}} & \multicolumn{2}{r}{\textbf{XGBoost}} & \multicolumn{2}{r}{\textbf{TiDE}} \\
\cmidrule(lr){3-4}\cmidrule(lr){5-6}\cmidrule(lr){7-8}
 &  & MAE & RMSE & MAE & RMSE & MAE & RMSE \\
\midrule
None & 0.000 & 0.021 & 0.030 & 0.029 & 0.041 & 0.028 & 0.039 \\
\midrule
bior1.1 & 0.400 & 0.021 & 0.030 & 0.029 & 0.041 & 0.036 & 0.050 \\
bior1.1 & 0.600 & 0.021 & 0.030 & 0.030 & 0.043 & 0.027 & 0.052 \\
bior1.1 & 0.800 & 0.022 & 0.031 & 0.028 & 0.040 & 0.026 & 0.037 \\
bior1.1 & 0.900 & 0.022 & 0.032 & 0.030 & 0.043 & 0.038 & 0.052 \\
bior1.1 & 0.950 & 0.023 & 0.034 & 0.030 & 0.043 & 0.029 & 0.043 \\
bior1.1 & 0.990 & 0.026 & 0.037 & 0.033 & 0.046 & 0.038 & 0.053 \\
bior1.1 & 0.999 & 0.035 & 0.048 & 0.068 & 0.095 & 0.062 & 0.092 \\
\midrule
bior1.5 & 0.400 & 0.021 & 0.030 & 0.030 & 0.043 & 0.022 & 0.032 \\
bior1.5 & 0.600 & 0.021 & 0.030 & 0.030 & 0.043 & 0.032 & 0.047 \\
bior1.5 & 0.800 & 0.022 & 0.031 & 0.029 & 0.042 & 0.075 & 0.098 \\
bior1.5 & 0.900 & 0.022 & 0.032 & 0.031 & 0.044 & 0.071 & 0.087 \\
bior1.5 & 0.950 & 0.023 & 0.033 & 0.031 & 0.045 & 0.047 & 0.063 \\
bior1.5 & 0.990 & 0.027 & 0.038 & 0.037 & 0.050 & 0.030 & 0.042 \\
bior1.5 & 0.999 & 0.052 & 0.072 & 0.371 & 0.402 & 1.356 & 1.767 \\
\midrule
bior2.8 & 0.400 & 0.021 & 0.030 & 0.030 & 0.043 & 0.024 & 0.034 \\
bior2.8 & 0.600 & 0.079 & 0.123 & 0.029 & 0.042 & 0.036 & 0.051 \\
bior2.8 & 0.800 & 0.138 & 0.225 & 0.030 & 0.045 & 0.058 & 0.075 \\
bior2.8 & 0.900 & 0.172 & 0.273 & 0.031 & 0.046 & 0.024 & 0.035 \\
bior2.8 & 0.950 & 0.219 & 0.383 & 0.031 & 0.045 & 0.058 & 0.075 \\
bior2.8 & 0.990 & 0.381 & 0.842 & 0.047 & 0.065 & 0.060 & 0.083 \\
bior2.8 & 0.999 & 0.489 & 0.974 & 0.373 & 0.410 & 1.895 & 2.621 \\
\midrule
bior3.9 & 0.400 & 0.021 & 0.030 & 0.030 & 0.043 & 0.048 & 0.066 \\
bior3.9 & 0.600 & 8.207 & 13.049 & 0.031 & 0.044 & 0.046 & 0.062 \\
bior3.9 & 0.800 & 11.783 & 18.837 & 0.031 & 0.044 & 0.021 & 0.030 \\
bior3.9 & 0.900 & 20.217 & 33.135 & 0.031 & 0.045 & 0.082 & 0.105 \\
bior3.9 & 0.950 & 27.882 & 48.939 & 0.031 & 0.044 & 0.053 & 0.072 \\
bior3.9 & 0.990 & 37.194 & 63.183 & 0.051 & 0.071 & 0.073 & 0.110 \\
bior3.9 & 0.999 & 57.729 & 118.634 & 0.274 & 0.314 & 2.744 & 9.379 \\
\midrule
bior6.8 & 0.400 & 0.021 & 0.030 & 0.030 & 0.042 & 0.027 & 0.038 \\
bior6.8 & 0.600 & 4197.230 & 6372.656 & 0.030 & 0.044 & 0.037 & 0.052 \\
bior6.8 & 0.800 & 5700.958 & 9089.746 & 0.030 & 0.044 & 0.071 & 0.093 \\
bior6.8 & 0.900 & 6617.189 & 10744.732 & 0.031 & 0.045 & 0.027 & 0.038 \\
bior6.8 & 0.950 & 8017.490 & 13269.162 & 0.032 & 0.045 & 0.037 & 0.051 \\
bior6.8 & 0.990 & 8729.190 & 14148.763 & 0.047 & 0.065 & 0.099 & 0.174 \\
bior6.8 & 0.999 & 37111.153 & 63630.100 & 0.309 & 0.348 & 5.112 & 12.458 \\
\bottomrule
\end{tabular}
\end{table}